\documentclass{aa}
\usepackage{times}

\newcommand{\ltsima} {$\; \buildrel < \over \sim \;$}
\newcommand{\simlt}  {\lower.5ex\hbox{\ltsima}}            
\newcommand{\gtsima} {$\; \buildrel > \over \sim \;$}
\newcommand{\simgt}  {\lower.5ex\hbox{\gtsima}}            

\newcommand{\be} {\begin{equation}}

\newcommand{\ee} {\end{equation}}

\newcommand{\bc}{\begin{center}}
\newcommand{\ec}{\end{center}}

\def \hcm {\hbox {\ifmmode $ atoms cm$^{-2}\else atoms cm$^{-2}$\fi}}
\def\deg {^\circ}

\def\ee {G21.5--0.9~}

\begin{document}


\title
{Timing analysis of the core of the Crab-like SNR \ee}

\author{N. La Palombara \& S. Mereghetti}

\institute {
{Istituto di Fisica Cosmica ``G.Occhialini'',
via Bassini 15, I-20133 Milano, Italy}
}

\offprints{N.La Palombara, nicola@ifctr.mi.cnr.it}

\date{Received / Accepted}

\authorrunning{N.La Palombara \& S.Mereghetti}
\titlerunning{Timing analysis of \ee}

\abstract{The Crab-like SNR G21.5-0.9 was observed in the X--ray
band (0.5-10 keV) by the {\it XMM-Newton} satellite for over 100
ks. The large effective area of the {\it EPIC} instrument has
allowed us to to perform a deep search for pulsations from the
central core of \ee. No pulsations were found with upper limits
on the pulsed fraction  between 7.5 \% and 40 \% (depending on
frequency and energy range).
  }

\maketitle

\section{Introduction}

Almost 5\% of the $\sim$ 225 known supernova remnants (SNRs) in our galaxy are
classified as ``Crab-like'' or ``plerionic'' (Green 2000).
From the spatial point of view, they are characterized by compact, centre-filled radio
and X--ray morphology (Weiler \& Panagia 1978).
These SNRs show featureless power-law spectra,
with a relatively flat spectral index in the radio regime
($\alpha_{r}\sim$0.0 - 0.3) and a steeper one at shorter wavelengths,
which are typical of synchrotron processes.

These morphological and spectral characteristics are explained by the presence of a central pulsar,
which injects high energy electrons that suffer synchrotron radiation losses as they diffuse through the surrounding
magnetic field  (see, e.g., Reynolds \& Chanan 1984).

\ee shows many characteristics of the Crab-like remnants.
Both in the radio band (Morsi \& Reich 1987)
and at X--ray energies (Becker \& Szymkowiak 1981) its emission is centrally peaked.
Evidence for a non-thermal X--ray spectrum was also indicated by {\it GINGA}
observations (Asaoka \& Koyama 1990).

The available measurements of the neutral hydrogen absorption give a distance
of $\sim$ 4.8 kpc (Becker \& Szymkowiak 1981).
The radio luminosity of \ee is $\sim$ 1.8$\times$10$^{34}$ {\it d}$^{2}_{5}$  erg s$^{-1}$ (Morsi \& Reich 1987),
i.e. a factor $\sim$ 9 smaller  than that of the Crab, but its X--ray luminosity is a factor $\sim$ 100 less,
therefore the {\it L$_{X}$/L$_{r}$} ratio is significantly lower.

X--ray observations performed with the \textit{Chandra} satellite detected a compact central core
of $\sim$ 2$''$ in size, at the center of the more  extended synchrotron nebula
of $\sim$ 30$''$ radius (Slane et al. 2000).
The central core, which is spatially resolved, most likely
marks the position of the pulsar powering \ee.
Also a fainter, more extended ``halo'' (radius $\sim 2'$) was detected with \textit{Chandra}.
This was tentatively interpreted as  the outer ``shell'' formed by the expanding ejecta
and the passage of the supernova-driven blastwave (Slane et al. 2000).
However, a   recent {\it XMM-Newton} observation  shows that also the halo
has a non-thermal spectrum;
it is probably a low surface brightness extension of the plerionic nebula
(Warwick et al. 2001).
However, this interpretation is not supported by radio data, since no
significant   radio emission
has yet been detected at such a large distance  from the source core (Bock \& Wright 2001).

Up to now no pulsed emission has been detected in the radio or X--ray energy range from the
putative neutron star at the center of \ee
(Frail \& Moffet 1993; Kaspi et al. 1996; Biggs \& Lyne 1996; Slane et al. 2000).
In this paper we report the results of a sensitive timing analysis on data provided
by four {\it XMM-Newton} observations.

\section{Observations and data analysis}

During April 2000, \ee was observed four times by the {\it XMM-Newton} mission
as one of the calibration targets.
A log of the observations is given in Table~1.
The source was on-axis in the first observation and $\sim$ 10$'$ off-axis in the following
ones; in each case the accumulated exposure time was $\sim$ 30 ks.
The results reported by Warwick et al. (2001) were based only on the on-axis
observation.

\begin{table*}[tbp]
\begin{center}
  \caption{Summary of the observations}
  \begin{tabular}[c]{|c|c|c|c|c|}
\hline
Start time (UT)  & Off-axis angle        & Exposure      & MOS frame     & PN frame \\
April 2000      & arcmin                & ks            & s             & ms \\
\hline
7, 12h 35m    & 0.2           & 30    & 2.6   & 200 \\
9, 12h 22m    & 10.3          & 29    & 2.6   & 73 \\
11, 12h 26m   & 10.4          & 29    & 2.6   & 73 \\
15, 12h 26m   & 10.2          & 29    & 2.6   & 73 \\
\hline
\end{tabular}
\end{center}
\end{table*}

In all the observations the source was imaged by the three focal plane {\it CCD} cameras
({\it MOS1}, {\it MOS2} and {\it PN}) of the {\it EPIC} instrument
(Turner et al. 2001, Strueder et al. 2001).
Each of the {\it MOS} cameras provides an effective area of $\sim$600 cm$^2$ at 1.5 keV,
and covers the energy range 0.2-10 keV;
for the {\it PN} camera the corresponding values are, respectively, $\sim$1400 cm$^2$
and 0.15-15 keV.
All the instruments were operated with the medium filter.
The {\it PN} camera worked in {\it Extended Full Frame}  mode in the first
observations, with a {\it CCD} frame time of 200 ms, and in standard {\it Full Frame}
mode in the following ones, with a frame time of 73 ms;
both {\it MOS} cameras used the  {\it Full Frame}  mode in all the observations,
with a frame time of 2.6 s.

The first step of our data analysis was the event selection.
For each of the four observations and for each of the three cameras we extracted all the events,
with energy between 1 and 10 keV, from a circular region with  radius of 25$''$
centered at the peak of the X--ray emission
(RA = 18h 33m 33.8s, DEC = -10$\deg$ 34$'$ 6$''$ (J2000)).
This radius contains $\sim$80\% of the photons for a point source.
We only considered events with pattern in the range 0-12 and 0-4 for, respectively,
the {\it MOS} and the {\it PN} camera, in order to reject the non X--ray events
due to cosmic rays and cosmetic defects.

The times of the selected events were converted to the   Solar System barycenter
with the {\it Reconstructed Orbit Files} provided by the {\it XMM Survey Science Center}.
These events were still tagged with discrete arrival times,
corresponding to the readout times of the individual CCD frames:
these arrival times were ``randomized'' by subtracting a random value between 0 and the
relevant frame time from the original times.

On these data we performed search for periodicities based on Fourier analysis
(see, e.g., van  der Klis 1989).
A Fourier power spectrum was computed for each of the instruments and observations and
examined for the presence of peaks above a threshold corresponding to a chance
probability of 10$^{-3}$ of being exceeded in the absence of a signal in a single spectrum.
No significant peaks were found.

To increase the detection sensitivity we repeated the same analysis on the sum of the
individual spectra.
To this aim, due to the three different  CCD frame times (0.073, 0.2 and 2.6 s),
we had to consider three different frequency ranges for the data analysis (Tab.2):
\begin{itemize}
\item{for $\nu <$ 0.1923 Hz = (2$\times$2.6 s)$^{-1}$, we summed the power spectra of all
      the observations and instruments}.
\item{for 0.1923 Hz $< \nu <$ 2.5 Hz = (2$\times$0.2 s)$^{-1}$, we summed the power spectra
      of only the {\it PN} data of the four observations}.
\item{for 2.5 Hz $< \nu <$ 6.85 Hz = (2$\times$0.073 s)$^{-1}$, we summed the power spectra of the
      {\it PN} data of the three off-axis observations}.
\end{itemize}
Again, no statistically significant peak was found in the summed power spectra.
We also repeated the whole procedure for two
distinct energy ranges, 1-3.5 keV  and 3.5-10 keV, again without any significant pulsed
signal.

To compute the upper limits on the source pulsed fraction,
in the assumption of a sinusoidal pulse shape,
we followed the procedure described in  van der Klis (1989), taking into account the
relevant correction pointed out by  Vaughan et al. (1994).
Finally, we had to correct the resulting upper limits to take into account the fraction
of unpulsed flux due to the nebular emission.
We based this correction on the results of the
\textit{Chandra} observation (Slane et al. 2000),
showing that at least 92\% of the flux within 30$''$ is of diffuse origin.
Thus we obtained the upper limits (99.9 \% confidence level) on the pulsed fraction of the
neutron star emission given in Table~2, where we also report the number of counts used in the
analysis.

\begin{table*}[tbp]
\begin{center}
  \caption{Upper limits on the pulsed fraction (99.9 \% c.l.)}
  \begin{tabular}[c]{|c|c|c|c|c|c|c|}
\hline
Frequency range & \multicolumn{2}{c}{1-10 keV}      & \multicolumn{2}{|c}{1-3.5 keV}     & \multicolumn{2}{|c|}{3.5-10 keV}  \\
Hz              & \%     & counts       & \%      & counts      & \%  & counts \\
\hline
$<$ 0.19        & 6.2-7.5      & 522142     & 9.1-11.0    &   300332  & 12.4-13.5 & 221810  \\
0.19 $<$ 2.5    & 15.1-17.1  & 275543     & 19.5-22.1    & 154231   & 24.2-27.3 & 121312  \\
2.5 $<$ 6.8     & 24.6-30.1   & 170434     & 32-39.2      & 98105   & 33.3-40.8 & 72329  \\
\hline
\end{tabular}
\end{center}
\end{table*}

\section{Discussion}


Our upper limits can
be compared with those obtained by some recent works based on X--ray
data of \ee.
 Note that the values reported by Warwick et al. (2001),
3.5\% and 5.5\% (respectively for MOS+PN and  PN only data), refer to the \textit{total}
flux within an extraction radius of 8$''$, including the nebular emission.
Our results for the on-axis observation alone (the one used by these authors) are
similar, but we reached a better sensitivity  in the summed power spectra
(corresponding to a factor $\sim$4 greater exposure time).

For instrumental reasons, our  search was limited to periods longer than 146 ms.
Although it is clearly possible that the pulsar in G21.5-0.9 has a
shorter period, we note that Safi-Harb et al. (2001), 
from energetic
considerations, estimated a period of P = 0.144 $(I_{45} / (\dot E_{37} \tau_{3}))^{1/2}$ s
(where $I=10^{45}I_{45}$ g cm$^{2}$ is the moment of inertia, $\dot E = 10^{37} \dot E_{37}$ 
erg s$^{-1}$ is the spin-down energy loss and $\tau = \tau_{3}\times$(3 kyr) is the pulsar characteristic
age). For reasonable values of $\tau$ and $\dot E$, such a period falls in the range we could explore
with EPIC. The same authors analyzed  five data sets (total 75 ks) obtained with the
\textit{Chandra HRC} instrument. They report an upper limit of 16\%, without quoting the
confidence level and, presumably, referring to the total flux within 2$''$.

The ``canonical'' picture of   plerionic supernova remnants is based on young,
energetic neutron stars with short rotation periods, such as the  Crab pulsar
(P=33 ms) or the recently discovered pulsar in 3C 58 (P=66 ms, Murray et al. 2001).
However, other results   show that also sensitive searches for slower pulsars
are relevant: there are in fact relatively young pulsars with long periods.
Besides the well known  example of  PSR B1509--58 in the SNR G320.4-01.2 (P=150 ms),
other recent findings include the 325 ms pulsar in the SNR Kes 75 (Gotthelf et al. 2000),
and PSR J1119--6127 (P=407 ms, Pivovaroff et al. 2001), which however does not have
a bright synchrotron nebula.





\newpage

\section{Acknowledgements}

We wish to thank the Italian Space Agency (ASI) for its financial support to this work.


\begin{thebibliography}{}

\bibitem{}  Asaoka I. \& Koyama K. 1990, PASJ 42, 625
\bibitem{}  Becker R.H. \& Szymkowiak A.E. 1981, ApJ 248, L23
\bibitem{}  Biggs J.D. \& Lyne A.G. 1996, MNRAS 282, 691
\bibitem{}  Bock D.C.J., Wright M.C. \& Dickel J.R. 2001, ApJ 561, L203
\bibitem{}  Frail D.A. \& Moffett D.A. 1993, ApJ 408, 637
\bibitem{}  Gotthelf E.V. et al. 2000, ApJ 542, 37
\bibitem{}  Green D.A. 2000, A Catalogue of Galactic Supernova Remnants, Cambridge UK, available at http://www.mrao.cam.ac.uk/surveys/snrs/
\bibitem{}  Kaspi V.M. et al. 1996, AJ 111, 2028
\bibitem{}  Morsi H.W. \& Reich W. 1987, A\&AS 69, 533
\bibitem{}  Murray S.S. et al. 2001, astro-ph/0108489
\bibitem{}  Pivovaroff M.J. et al. 2001, ApJ 554, 161
\bibitem{}  Reynolds S.P. \& Chanan G.A. 1984, ApJ 281, 673
\bibitem{}  Safi-Harb S. et al. 2001, ApJ 561, 308
\bibitem{}  Slane P. et al. 2000, ApJ 533, L29
\bibitem{}  Strueder L. et al. 2001, A\&A 365, L18
\bibitem{}  Turner M.J.L. et al. 2001, A\&A 365, L27
\bibitem{}  van der Klis M. 1989, in ``Timing Neutron Stars'', NATO ASI Ser.C, ed. H. \"Ogelman \& E.P. van den Heuvel (Dordrecht: Kluwer), 262, 27
\bibitem{}  Vaughan B.A. et al. 1994, ApJ 435, 362
\bibitem{}  Warwick R.S. et al. 2001, A\&A 365, L248
\bibitem{}  Weiler K.W. \& Panagia N. 1978, A\&A 70, 419

\end{thebibliography}
\end{document}